\documentclass[12pt]{article}
\usepackage[centertags]{amsmath}
\usepackage{footnote}
\usepackage[flushleft]{threeparttable}
\usepackage{amsfonts}
\usepackage{amssymb}
\usepackage{amsthm}
\usepackage{amsmath}
\usepackage{newlfont}
\usepackage{graphicx}
\usepackage{subfig}
\usepackage{placeins}
\usepackage{caption}
\usepackage{booktabs}
\usepackage{amsthm}
\usepackage{hyperref}
\usepackage{epsfig}
\usepackage{color}
\usepackage{latexsym}
\usepackage{graphicx}
\usepackage{booktabs}
\usepackage{array}
\usepackage{rotating}
\usepackage{xcolor}
\usepackage[table]{xcolor}
\newfont{\bb}{msbm10 at 14pt}

\usepackage{setspace}

\usepackage[left=2cm,right=2cm,top=2.3cm,bottom=2.3cm]{geometry}
\usepackage[font=small,skip=0.1pt]{caption}
\usepackage[T1]{fontenc}
\usepackage[utf8]{inputenc}
\usepackage{authblk}
\usepackage{booktabs}
\usepackage{siunitx}
\usepackage{multirow}
\date{}

\begin{document}
	\begin{center}{\Large\bf{ The Effect of Topological Defects and Magnetic Flux on Tetraquarks Using the Analytical Exact Iteration Method }}
	\end{center}
	\begin{center}
	\def\baselinestretch{1}
	N. H. Gerish\footnote{norhangerish@gmail.com}, M. Abu-Shady \footnote{dr.abushady@gmail.com } 
	and E. M. Khokha   \footnote{emad.khokha@ksiu.edu.eg } 
\end{center}
\begin{center}
	\small
	{$^{1}$ Department of Mathematics and Computer Science, Faculty of Science, Suez Canal University, Egypt.\\
		$^{2}$ Department of Mathematics and Computer Science, Faculty of Science, Menoufia University,  Egypt.\\
		$^{3}$	Faculty of Computer Science and Engineering, King Salman International University (KSIU), South Sinai,  Egypt}
\end{center}
	
	\begin{abstract}
	Investigating the non-perturbative behavior of QCD and the dynamics of strong interaction is crucial for the study of heavy quarkonia and the understanding of exotic fully-heavy tetraquarks. In this work, using the analytical exact iteration method (AEIM), the analytical eigenvalue solutions of the non-relativistic Schrödinger equation are obtained in the presence of topological defects and external magnetic flux. The interactions are modelled using a modified Cornell potential supplemented by harmonic and inverse quadratic terms. We demonstrate that the energy levels are distinctly shifted by the topological defect parameter ($\alpha$). The mass spectra of heavy quarkonia  ($c\bar{c}$ and $b\bar{b}$)  and fully-heavy tetraquarks ($cc\bar{c}\bar{c}$ and $bb\bar{b}\bar{b}$)  across several radial and orbital excitation states are successfully calculated using this approach. The computed masses of bottomonium and charmonium accord well with current theoretical predictions and experimental findings. Our findings for the heavy tetraquarks are in line with previous theoretical investigations that consider tetraquarks as configurations of diquarks and antidiquarks. The numerical results demonstrate that a nontrivial interaction between the confining potential and the background space-time geometry governs the mass hierarchy of these exotic hadronic states, providing high-precision data with excellent agreement with established theoretical models and experimental benchmarks.		
\end{abstract}
	\section{Introduction}
One of the most active areas of modern hadron physics is the study of exotic hadrons, or states whose quark configurations go beyond the standard meson and baryon \textit{qqq} image. A wide range of tetraquark and pentaquark possibilities have been found since the Belle Collaboration discovered the X(3872) in 2003, especially in the heavy-flavour sector. A very clean laboratory for studying quantum chromodynamics (QCD) is provided by fully-heavy tetraquarks made entirely of charm and/or bottom quarks. These systems offer a direct probe of the confining force and the non-Abelian nature of colour
interactions since they are not subject to the intricacies of light-quark dynamics and pion-exchange interactions, in contrast to their light or heavy-light counterparts \cite{j18}-\cite{f1}. However, there is still uncertainty regarding the theoretical status of fully heavy tetraquarks. Using rigorous four-body calculations within constituent quark models, Richard, Valcarce et al. \cite{j18} showed that equal-mass configurations such ($cc\bar{c}\bar{c}$ and $bb\bar{b}\bar{b}$) are not bound under exclusively chromoelectric potentials. On the other hand, string-inspired multibody confinement processes may make the mixed-flavor configuration metastable, emphasizing the crucial significance of color algebra and quark-mass asymmetry. This result emphasises how tetraquark binding is determined by a tight balance between kinetic energy, confinement, and chromomagnetic interactions.\\
The spectroscopy of heavy tetraquarks and pentaquarks has been improved by parallel theoretical research using two new components: topological defect space–times and fractional-order dynamics.   In a recent study, Ongodo et al. \cite{ph23} used generalized fractional derivatives in conjunction with a point-like global monopole (PGM) background to study the hyperfine mass splittings of the ground and radially excited states of heavy-flavoured tetraquarks. Their findings show that both fully-charmed and fully-bottom tetraquarks mass spectra are sensitive to the fractional order and the topological defect parameter. Importantly, it was discovered that some radially excited states were below the appropriate two-meson thresholds, indicating potential metastability against fall-apart decays, a crucial requirement for experimental observability. In a similar vein, Atangana et al. \cite{f2} extended this approach to heavy pentaquarks, yielding ground-state masses for configurations with spin-parities in a fractal space-time with a PGM. For small values of the topological defect parameter, the results showed congruence with available experimental evidence. The impact of topological defects on heavy mesons in hot and isotropic media was examined by the authors in Refs. \cite{ph25}-\cite{h14}. \\
In particle physics, the hunt for magnetic monopoles is one of the oldest and most significant endeavors outside of the Standard Model. The existence of even a single magnetic monopole would explain the quantization of electric charge and restore electromagnetic duality to Maxwell's equations, according to Dirac's well-known argument. Using the Schwinger mechanism as the production pathway, the ATLAS Collaboration recently reported a search for magnetic monopole pair creation in ultraperipheral Pb+Pb collisions \cite{h15}. Upper bounds on the production cross section for monopoles with a single Dirac magnetic charge in the mass range of $20-150$ GeV were obtained since no appreciable excess over the background was found. Monopole masses below $80-120$ GeV are omitted, depending on the theoretical model (locally constant field approximation or free particle approximation). These findings impose strict limitations on any situation involving light magnetic charges. Cosmic strings are the most obvious topological defects because their geometry is symmetrical except for the symmetry axis. They originated early in the universe's history. The variety of ideas produced by general relativity theory provides a strong incentive to investigate how particles interact with these geometrical structures, despite the lack of fundamental proof for their existence \cite{h15}. They do not, however, fully embrace the 4D tetraquark and pentaquark proposal due to its departure from mainstream quantum field theory and lack of experimental validation. They investigate whether the hyperfine splitting, stability, and decay patterns of exotic hadrons could be affected by effective magnetic-moment terms or confining pressures that imitate magnetic charges. Indirect constraints or signatures that could be tested at the LHC and future colliders could be obtained from such an examination \cite{h18}. The Cosmic Microwave Background, which has studied topological flaws in quantum systems using a point-like global monopole, may cause matter compression and temperature oscillations due to the presence of cosmic strings \cite{h20}. Hydrogen and pionic atoms \cite{h41}, quantum motion of a spin-zero particle with potential under the AB-flux field \cite{h42}, the Dirac equation \cite{h43}, and the DKP equation under the AB-flux field and Coulomb potential \cite{h44} are among the investigations conducted in the relativistic limit. Cornell-type potential \cite{h50}, Mie-type potential when the AB-flux field is present \cite{h51}, and when the AB-flux field with the diatomic molecular potential is studied in Ref. \cite{h54}.         
By critically analysing the spectroscopy of fully heavy tetraquarks and heavy mesons within a unified framework that includes a rigorous treatment of the four-body system using the diquark-antidiquark iteration method, which is not considered in the previous works. However, the effects of topological defects and magnetic flux, we hope to bridge these disparate research works.
This is the structure of the paper. In Section 2, we solve the Schrödinger equation under a cosmic string background using the Cornell, harmonic, and inverse-harmonic potentials. The results for heavy diquark and tetraquark systems are discussed in Section 3. We make our conclusions in Section 4.     
\section{The Non-Relativistic Schrödinger Equation under Topological Defect and External Magnetic Flux  }
\begin{figure}[!h]
	\vspace{1.5cm}
	\centering
	\begin{minipage}{.5\textwidth}
		\centering
		\includegraphics[width=1\textwidth]{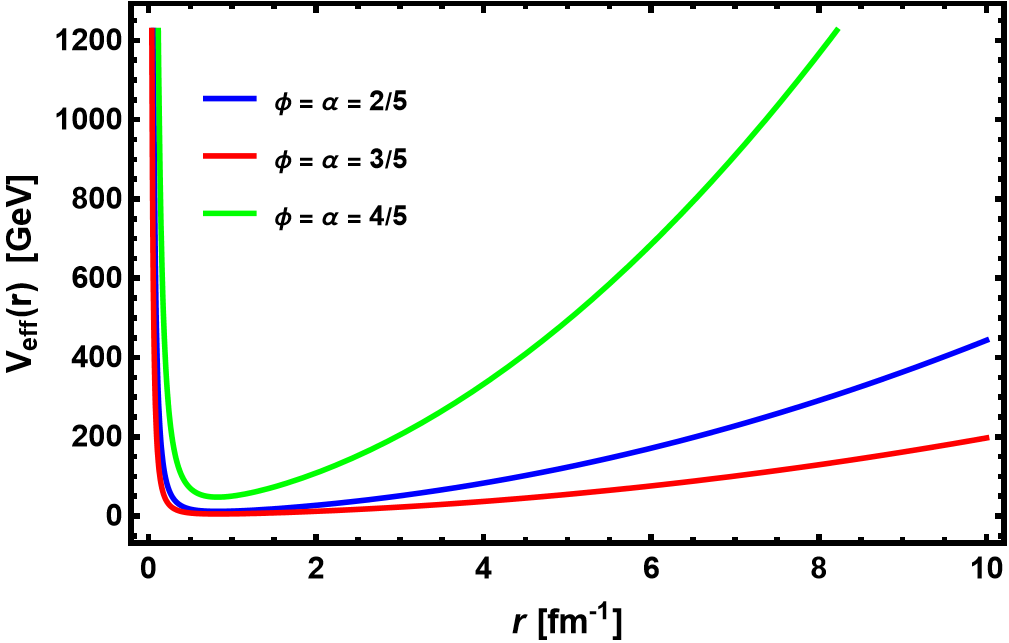}
		\textbf{(a) $a=b=c=d=m=l=1$}
	\end{minipage}%
	\begin{minipage}{.5\textwidth}
		\centering
		\includegraphics[width=1\textwidth]{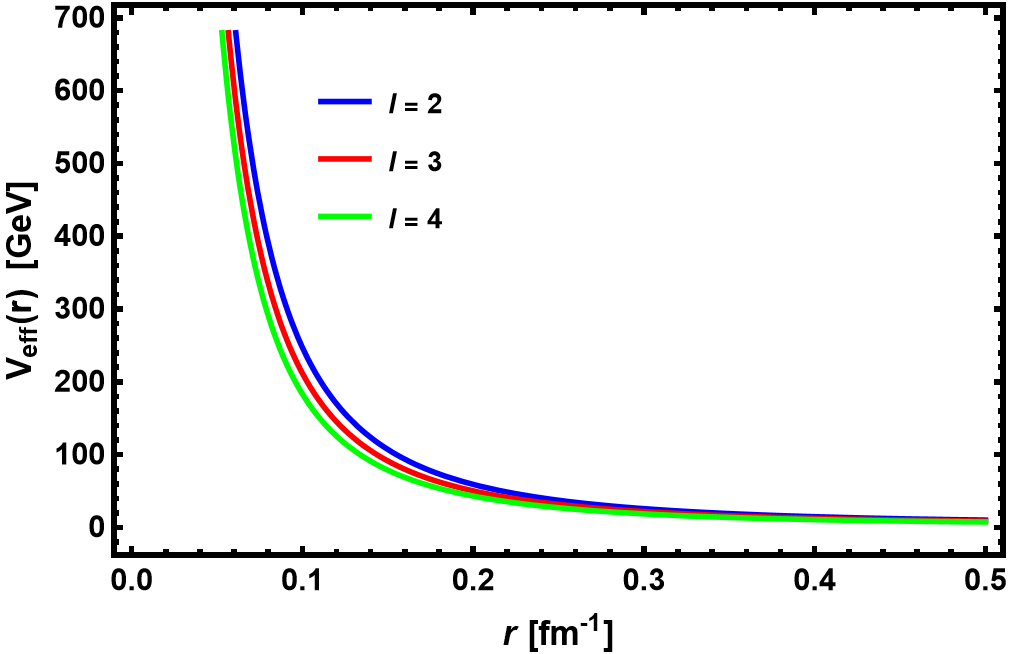}	
		\textbf{(b) $a=b=c=d=1,\phi=1 $ and $\alpha=\frac{3}{4}$}
	\end{minipage}%
	
	\begin{minipage}{.5\textwidth}
		\centering
		\includegraphics[width=1\textwidth]{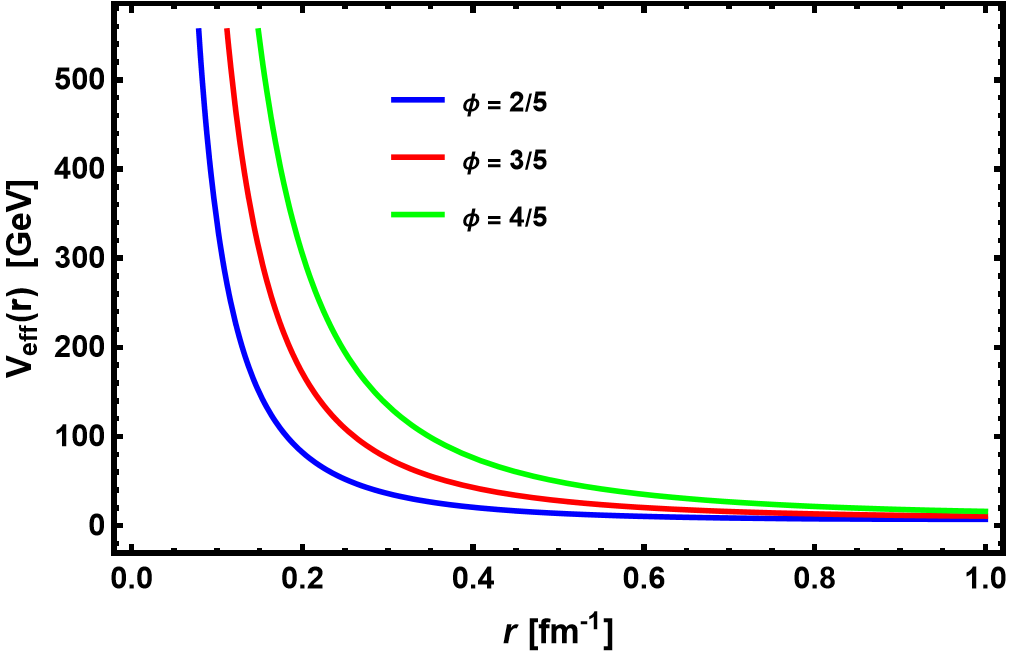}
		\textbf{(c) $a=b=c=d=1, l=1 $ and $\alpha=\frac{3}{4}$}
	\end{minipage}%
	\begin{minipage}{.5\textwidth}
		\centering
		\includegraphics[width=1\textwidth]{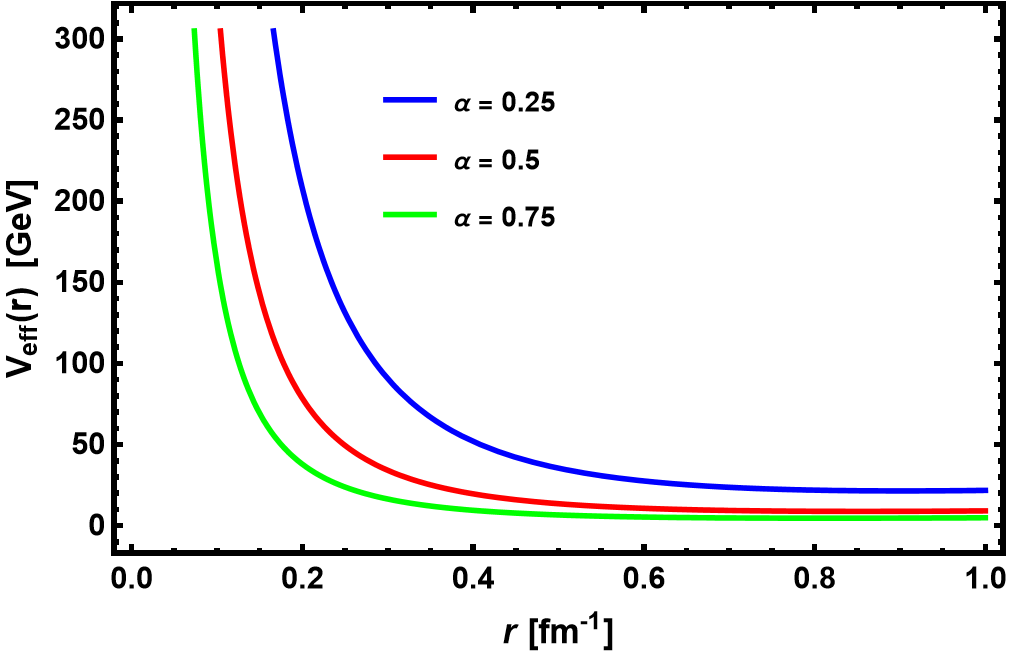}	
		\textbf{ (d)  $a=b=c=d=1, l=1  $ and $\phi=\frac{3}{4}$}
	\end{minipage}%
	\\
	\caption{\small{The behavior of the effective potential $V_{eff}(r)$ as a function of distance $r$, $(a)$ variation of $V_{eff}(r)$ with the magnetic flux and topological defect parameter $( \phi, \alpha)$. $(b)$ the effect of different orbital angular momentum $(l)$. $(c)$ and $(d)$ show the sensitivity of the potential to variations in the coupling parameters and $\alpha$ under specific constraints.   }}\label{h1}
\end{figure}

\begin{figure}[!h]
	\vspace{.5cm}
	\centering
	\begin{minipage}{.5\textwidth}
		\centering
		\includegraphics[width=1\textwidth]{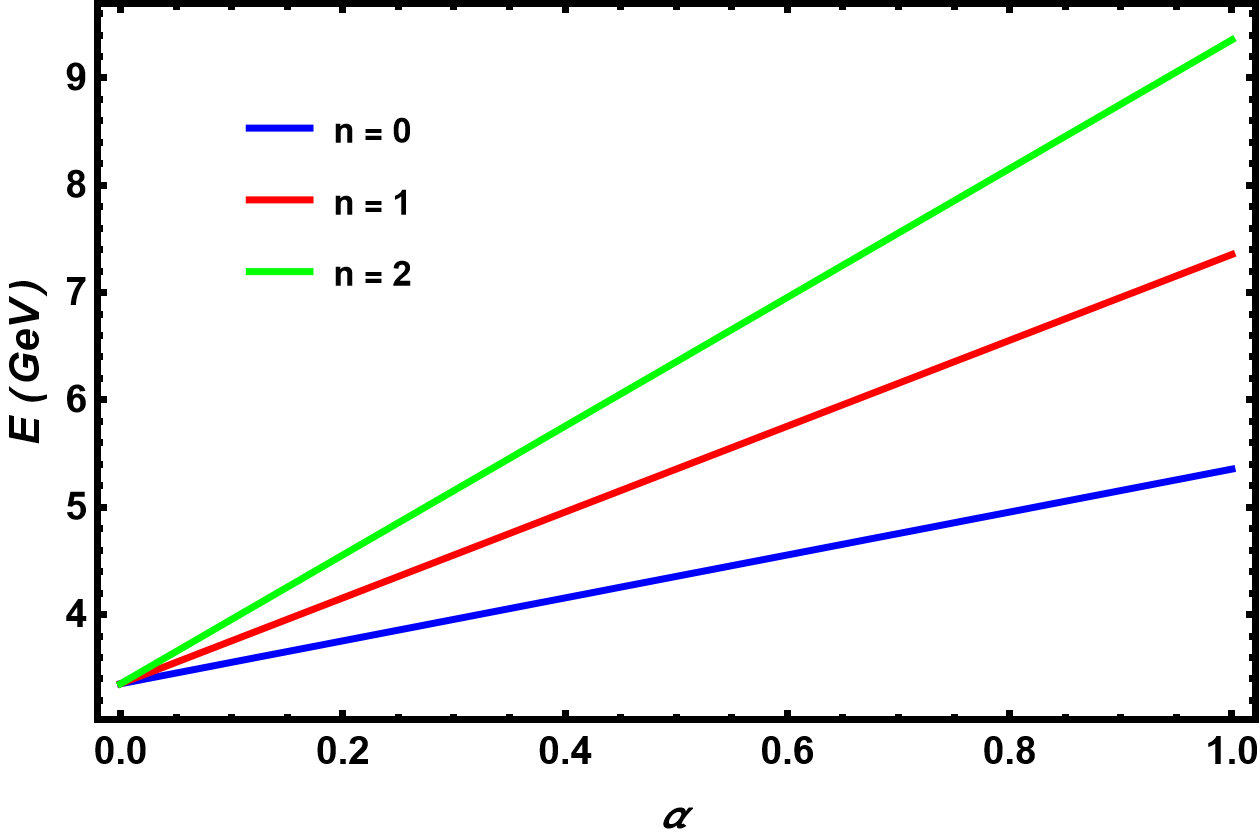}
		\textbf{(a) $a=b=c=d=m=1=l$ and $\phi=0$}
	\end{minipage}%
	\begin{minipage}{.5\textwidth}
		\centering
		\includegraphics[width=1\textwidth]{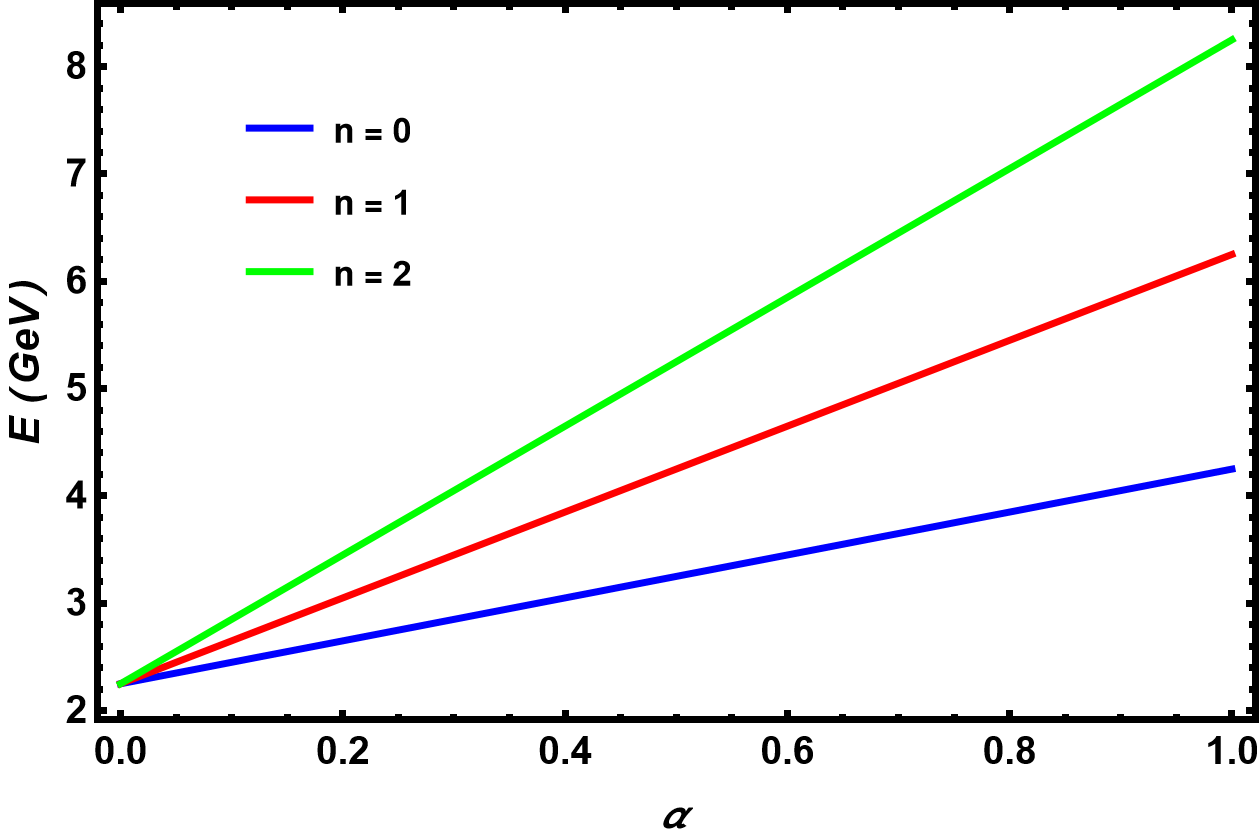}	
		\textbf{	(b) $a=b=c=d=l=1,$ and $\phi=\frac{3}{4}$ }
	\end{minipage}%
	
	\begin{minipage}{.5\textwidth}
		\centering
		\includegraphics[width=1\textwidth]{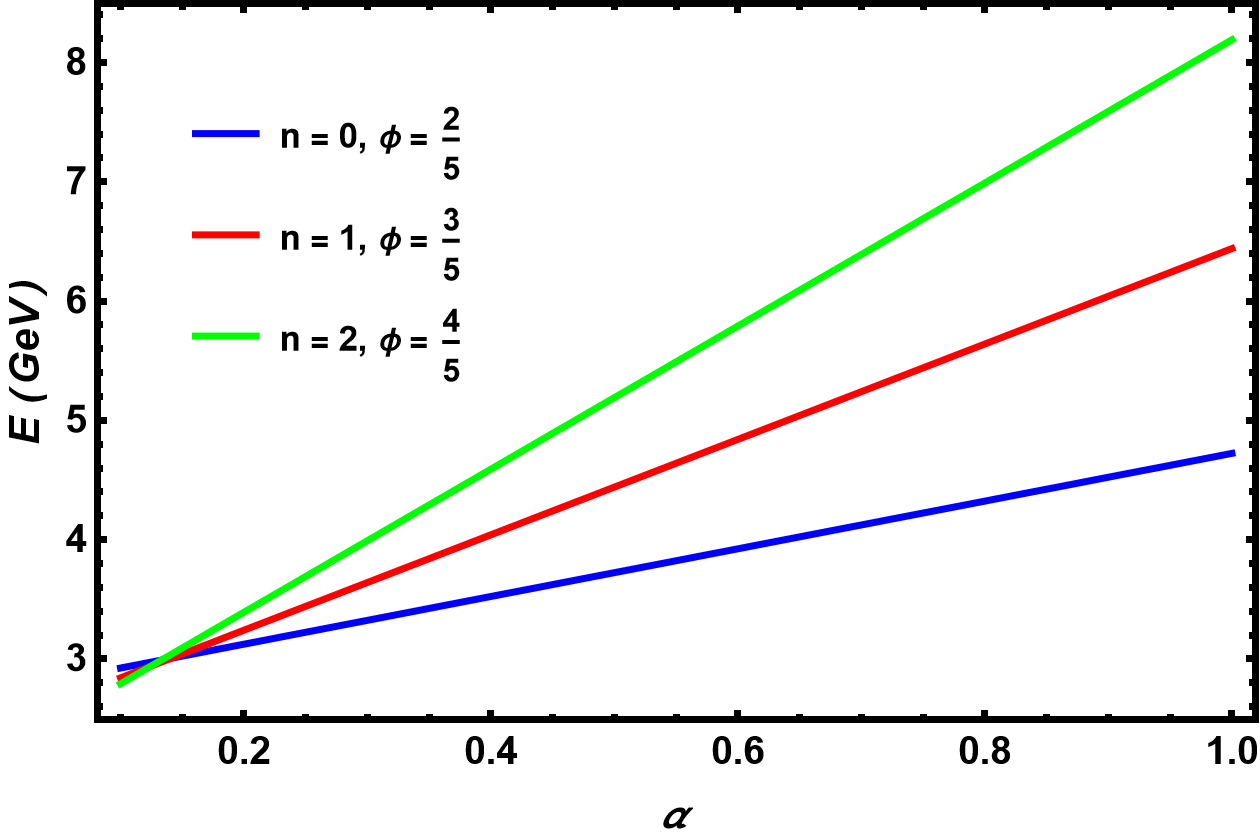}
		\textbf{(c)  $a=b=c=d=m=1,$ and $l=1$  }
	\end{minipage}%
	\begin{minipage}{.5\textwidth}
		\centering
		\includegraphics[width=1\textwidth]{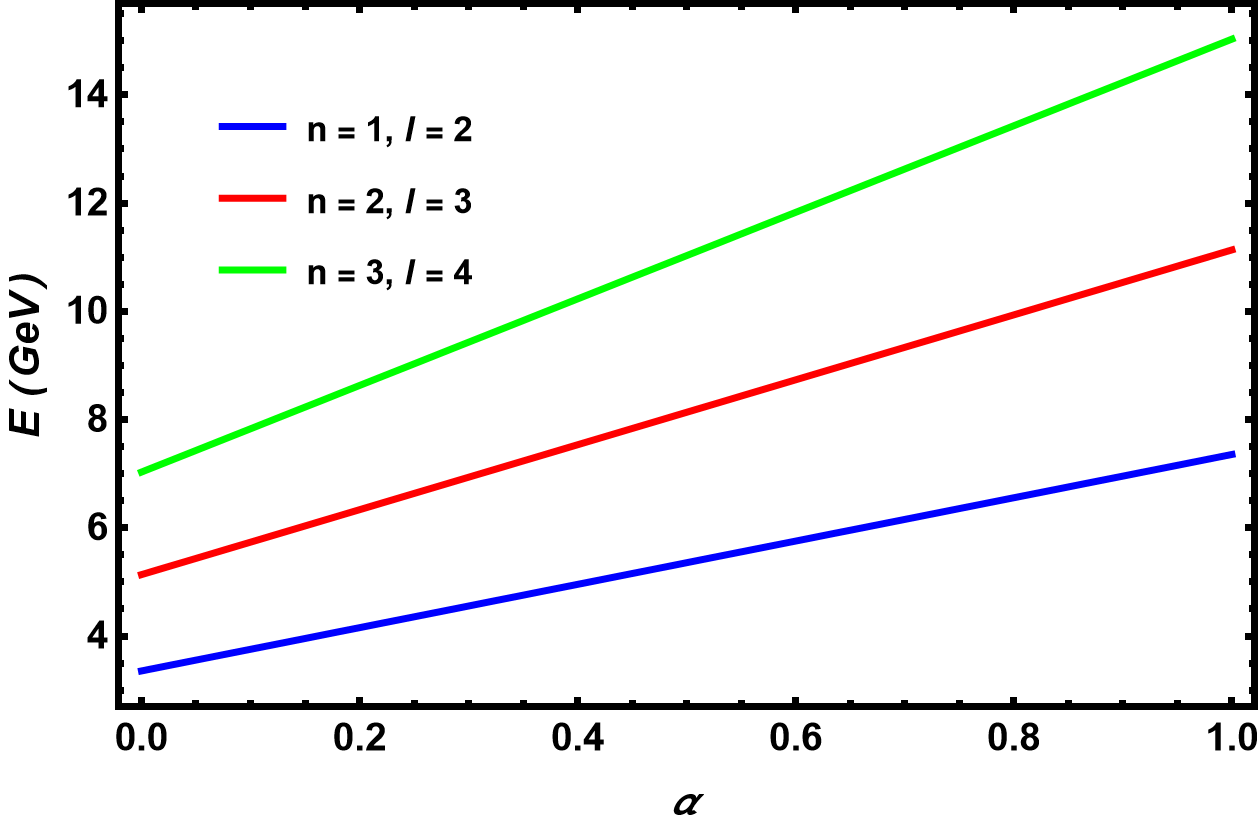}	
		\textbf{(d) $a=b=c=d=m=1,$  and $\phi=1$ }
	\end{minipage}%
	\\
	\caption{\small{The variation of energy eigenvalues E as a function of the topological defect parameter $\alpha$ for different quantum states. Subplots (a), (b), (c), and (d) illustrate the energy levels for varying principal quantum numbers (n), orbital angular momentum (l), and magnetic flux $(\Phi)$.  }}\label{h2}
\end{figure}

In the presence of an electromagnetic potential, the time-dependent Schrödinger wave equation is represented by the following equation:  \cite{h54, d53}
\begin{equation}\label{x1}
	\left[
	- \frac{1}{2\mu} 
	\left( \frac{1}{\sqrt{g}} D_i \sqrt{g} \, g^{ij} D_j \right) 
	+ V(r)
	\right] \Psi 
	= i \frac{d\Psi}{dt},
\end{equation} 
Where $\mu$ is the reduced mass of the system, $g$ is the determinant of the metric tensor $g_{ij}$, and $g^{ij}$ its inverse and the Gauge covariant derivative
$D_{i}= (\partial_{i}-i \, e \, A_{i}\,)$ since $e $ is the electric charges. The determinant of the spatial metric tensor $g_{ij}$ is under consideration for the space-time (\ref{x1}) is defined as $g= \frac{r^{2}\, \sin^{2}{\theta}}{\alpha^{2}}$.
\\
The total wave function $\Psi(t,r,\theta,\phi)$ has been defined in terms of several variables using the variable separation method.
Now we consider the following wave function:
\begin{equation}\label{x2}
	\Psi(t, r, \theta, \phi)= e^{i \,E \,t} \,\, Y_{l, m}(\theta, \phi) \psi_{\nu,\gamma}(r),
\end{equation}
Where $E$ the non-relativistic particle's energy, $ Y_{l,m}(\theta,\phi)$ is the spherical harmonic function, and $m, l$ are the magnetic moment and orbital quantum numbers, respectively.
The electromagnetic three-vector potential  $\vec{A}$ in this work is expressed as \cite{h54}:
\begin{equation}\label{x3}
	A_{r}=0,\,\,\ A_{\theta}=0,\,\,\,\  A_{\phi}=\frac{\Phi_{B}}{2\pi r \sin{\theta}},       \Phi_{B}=\Phi \, \Phi_{0},  \,\,\,\,\,\, \,\, \Phi_{0}= \frac{2\pi}{e},
\end{equation}
where $\Phi_{B}$ is a constant that represent the Aharonov-Bohm flux, $\Phi_{0}$ is the quantum of magnetic flux, and $\Phi$ is a real number which is the amount of magnetic flux.
\\
Thus, the wave equation (\ref{x1}) in the space-time has been expressed and used in Eqs. (\ref{x2}), (\ref{x3}). We acquire the following angular and radial differential equations for $ Y_{l,m} $, and $\psi(r)$, respectively as
\begin{equation}\label{x4}
	\frac{\alpha^{2}}{r^{2}} \frac{d }{dr} \Bigg(r^{2}\frac{d\psi_{\nu,\gamma}(r)}{dr}\Bigg)+ \Bigg[2\mu \,\,\Big(E-V(r)\,\Big)-\frac{l'(l'+1)}{r^{2}} \Bigg] \psi_{\nu,\gamma}(r)=0.
\end{equation}
and 
\begin{equation}\label{x5}
	\Bigg[\frac{1}{\sin{\theta}}\frac{d}{d\theta}\,+ \frac{1}{sin^{2}\theta} \Bigg(\frac{d}{d\phi}-i \Phi\Bigg)+l'(l'+1)\Bigg] Y_{l,m}(\theta,\phi)=0.
\end{equation}
where $l'= (l-\Phi)$ is the effective orbital quantum number and $\alpha$ is the topological defect parameter. The radial Schrödinger equation can be expressed as
\begin{equation}\label{x6}
	\psi''_{\nu,\gamma}(r)+\frac{2}{r}\psi'_{\nu,\gamma}(r)+ \frac{1}{\alpha^{2}} \Bigg[2\mu(E-V(r))-\frac{l'(l'+1)}{r^{2}}\Bigg] \psi_{\nu,\gamma}(r)=0.
\end{equation}
where $V_{eff}(r)$ is the effective potential, which plays a pivotal role in the analysis as it explicitly couples the confinement potential with the topological defect parameter $(\alpha)$ and the external magnetic flux $(\Phi)$. This expression demonstrates how the spacetime geometry and magnetic field collectively modify the centrifugal barrier, thereby shifting the energy eigenvalues of the diquark and tetraquark systems.
\begin{equation}\label{x7}
	V_{eff}(r)= \frac{1}{\alpha^{2}}\Bigg[V(r)+\frac{(l-\Phi) (l-\Phi+1)}{2\mu r^{2}}\Bigg].
\end{equation}
In this work, the confinement potential is $V(r)$ utilised to characterise quark-antiquark interactions. This model incorporates the Cornell potential (linear and Coulombic terms) to ensure color confinement at large distances and asymptotic freedom at short ranges. Additionally, harmonic and inverse-harmonic terms $(ar^2$ and $d/r^2)$ are included to account for the stability of the bound states and geometric corrections arising from the topological environment.
\begin{equation}\label{x8}
	V(r)=a r^{2} + b\, r- \frac{c}{r}+ \frac{d}{r^{2}}.
\end{equation}
The parameters $a, b, c,$ and $d$ in the potential $V(r)$ are assigned dimensions such that the total potential energy is expressed in GeV, consistent with the radial distance $r$ in GeV$^{-1}$ and we will determine them later.
By inserting the function $\psi_{\nu,\gamma}(r)=  r^{-1} R_{\nu,\gamma}(r)$ into the Eq. (\ref{x6}), we get:

\begin{equation}\label{x9}
	\frac{d^{2}R_{\nu,\gamma}(r)}{dr^{2}}+\Bigg[ \epsilon-a_{1}r^{2}-b_{1}r+\frac{c_{1}}{r}-\frac{d_{1}}{r^{2}}\Bigg]R_{\nu,\gamma}(r)=0.
\end{equation}
where 
\begin{equation}\label{x10}
	\epsilon=\frac{2\mu E}{\alpha^{2}},\,\,\,\,\,\,a_{1}=\frac{2\mu a}{\alpha^{2}} ,\,\,\,\,\,\,b_{1}=\frac{2\mu b}{\alpha^{2}},\,\,\,\,\,\,c_{1}=\frac{2\mu c}{\alpha^{2}}\,\,\,\,\,\,d_{1}=\frac{2\mu d+l'(l'+1) }{\alpha^{2}}.
\end{equation}
The wave function $R_{\nu,\gamma}(r)$ is defined as \cite{c1}:
\begin{equation}\label{x11}
	R_{\nu,\gamma}(r)=f_{\nu}(r)\cdot \exp \big[g_{\gamma}(r)\big].\end{equation}
and we take $ f_{\nu}(r)$, $ g_{\gamma}(r)$ as
\begin{equation}\label{x12}
	f_{\nu}(r)= \begin{cases} 
		1 &  \nu=0\\
		\prod_{i=1}^{\nu}(r-\alpha_{i}^{\nu}) &  \nu=1,2,..
\end{cases}\end{equation}
\begin{equation}\label{x13}
	g_{\gamma}(r)=-\frac{1}{2} \alpha' r^{2}-\beta r + \delta \ln{r},\qquad \alpha',\beta >0.
\end{equation}
and
\begin{equation}\label{x14}
	\frac{d^{2}R_{\nu,\gamma}(r)}{dr^{2}}=\Bigg[g''_{\gamma}+ g_{\gamma}^{\prime 2}+\Bigg(\frac{f''_{\nu}+2 f'_{\nu}g'_{\gamma}}{f_{\nu}}\Bigg)\Bigg]\,R_{\nu,\gamma}(r),
\end{equation}
where $f_{\nu}(r)$ are obviously equal to the Laguerre polynomials.
The comparison of Eq. (\ref{x9}) and (\ref{x14}) yields:

\begin{equation}\label{x15}
	g''_{\gamma}+g_{\gamma}^{\prime 2}+\Bigg(\frac{f''_{\nu}+2 f'_{\nu}g'_{\gamma}}{f_{\nu}}\Bigg)=-\epsilon+a_{1}r^{2}+b_{1}r-\frac{c_{1}}{r}+\frac{d_{1}}{r^{2}}.
\end{equation} 
When  $\nu=0 $, we have
\begin{equation}\label{x16}
	-\epsilon+a_{1}r^{2}+b_{1}r-\frac{c_{1}}{r}+\frac{d_{1}}{r^{2}}=-\acute{\alpha }-\frac{\delta}{r^{2}}+\Big(-\alpha' r-\beta +\frac{\delta}{r}\Big)^{2}.
\end{equation}
By equating the two sides of the power 
$r$ in Eq. (\ref{x16}) leads to the following expression for the energy and the corresponding constraint on the potential parameters:
\begin{equation}\label{x17}
	\begin{aligned}
		\epsilon= &\alpha ' (1+2\delta)-\beta^{2},\\
		a_{1}=&\alpha '^{2},       \,\,\,\,\,\,\,\,\,\,\,\,\,\,\,\, \alpha '=\sqrt{a_{1}},\\
		b_{1}=&2\alpha ',  \,\,\,\,\,\,\,\,\,\,\,\,\,\,\,\,\, \beta=\frac{b_{1}}{2\alpha '}\\
		c_{1}=&2\beta \delta,\\
		\delta=&\frac{1}{2}\Bigg(1\pm\sqrt{4d_{1}+1}\Bigg)=\frac{1}{2}\Big(1\pm h\Big), \,\,\,\,\,\,\,\,\,\,\, h=\sqrt{4d_{1}+1}.
	\end{aligned}
\end{equation}
By substituting the expression for $a_{1}$ and $b_{1}$ from Eq.(\ref{x10}) and $\alpha '$, $\beta$, and $\delta$ from Eq. (\ref{x17}), we obtain the ground state wave function and the corresponding ground state energy as follows:
\begin{equation}\label{x19}
	\psi_{0,\gamma}(r)=N\,\,r^{-1}\, r^{\frac{1+h}{2}} \,\,\exp\,\Bigg[-\sqrt{\frac{\mu a}{2}}r^{2}-\frac{2\mu c}{(1+h)}r\Bigg],
\end{equation}
\begin{equation}\label{x18}
	E_{0,\gamma}=\alpha \sqrt{\frac{a}{2\mu}}\Bigg(2+ \sqrt{\frac{ 8\mu d+ 4 (l-\Phi) (l-\Phi+1)+1}{\alpha}} \Bigg)-\sqrt{\frac{\mu}{2a}}\,b.
\end{equation}
For the first mode ($\nu=1$), we utilized $f_{\nu}(r)=(r-\alpha_{1}^{1})$ and $g_{\gamma}(r) $ that given in Eq. (\ref{x12}), (\ref{x13}) and substituted it into Eq. (\ref{x15}) yields:

\begin{equation}\label{x20}
	\begin{aligned}
		&-\alpha '\Big[1+2(\delta+1)\Big]+\alpha '^{2} r^{2}+\beta^{2}
		- 2\alpha ' \beta r + \frac{\delta(\delta-1)}{r^{2}}+ \frac{2\delta}{r(r-\alpha_{1}^{1})}-\frac{2\beta \delta}{r}-\frac{2(\alpha ' \alpha_{1}^{1}+\beta)}{(r-\alpha_{1}^{1})}=\\
		&-\epsilon+ a_{1}r^{2}+ b_{1}r- \frac{c_{1}}{r}+\frac{d_{1}}{r^{2}}.
	\end{aligned}
\end{equation}
The relations between coefficients $\alpha '$, $\beta$, $\delta$, and $\alpha_{1}^{1}$ and the potential parameters are computed as, 
\begin{equation}\label{x21}
	\begin{aligned}
		\epsilon= &\alpha '\Big( 1 + 2(\delta+1)\Big)-\beta^{2}\\
		a_{1}=&\alpha '^{2},\\
		b_{1}=&2\alpha ' \beta,\\
		c_{1}=&2\beta \delta + 2\alpha '  \alpha_{1}^{1} +2 \beta = \frac{2 \beta  \delta  \alpha_{1}^{1} + 2 \delta}{ \alpha_{1}^{1}},\\
		\delta=&\frac{1}{2}\Big(1\pm\sqrt{1+ 4 d_{1} }\Big)=\frac{1}{2}(1\pm h),\\
		\alpha_{1}^{1}=&\frac{1}{2\alpha '}\Big[-\beta \pm \sqrt{ \beta^{2} + 4 \alpha ' \delta}\Big].
	\end{aligned}
\end{equation}
The corresponding energy and wave function are:
\begin{equation}\label{x22}
	E_{1,\gamma}=\alpha \sqrt{\frac{a}{2\mu}}\Bigg(3+ \sqrt{ \frac{8\mu d+ 4 (l-\Phi) (l-\Phi+1)+1}{\alpha}}\Bigg) - \sqrt{\frac{\mu}{2a}}\,b,
\end{equation}
\begin{equation}\label{x23}
	\psi_{1,\gamma}(x)=N\,\,r^{-1} r^{\frac{1 + h}{2}} \Big(r-\alpha_{1}^{1}\Big)\,\,\exp\,\Bigg[-\sqrt{\frac{\mu a}{2}}r^{2}-\frac{2\mu c }{(1 + h)} r\Bigg].
\end{equation}
By applying this procedure for higher modes $(\nu=2,3,....)$, we obtain the general expressions for the energy and wave function at arbitrary $\nu$:
\begin{equation}\label{x24}
	E_{\nu,\gamma}=\alpha \sqrt{\frac{a}{2\mu}}\Bigg(2+ 2 \nu+ \sqrt{\frac{8\mu d+ 4 (l-\Phi) (l-\Phi+1)+1}{\alpha}}\Bigg) - \sqrt{\frac{\mu}{2a}}\,b,
\end{equation}
\begin{equation}\label{x25}
	\psi_{\nu,\gamma}(r)=N\,\,r^{-1} r^{\frac{1 + h}{2}} \prod_{i=1}^{\nu}\Big(r-\alpha_{i}^{\nu}\Big)\,\,\exp\,\Bigg[-\sqrt{\frac{ \mu  a}{2}} r^{2}-\frac{2 \mu  c}{(1+h)}r\Bigg].
\end{equation}
The numerical results and the behavior of the effective potential and the energy eigenvalues are graphically represented in Figs. (\ref{h1}- \ref{h2}) for various parameters.
	\section{ Results and Discussion}
We show in Fig. (\ref{h1}) that the interaction energy between quarks changes with distance $(r)$.
The potential is very high due to the centrifugal barrier at small distances, and it increases at large distances because of the Cornell and harmonic terms, which help maintain quark confinement.
In case of changing the parameter $\alpha$ shifts the curves, which changes the binding strength of the diquark.
Fig. (\ref{h2}), shows how the total energy ($E$) 
changes when we vary the space defect $(\alpha)$. 
The energy increases linearly as the topological 
defect parameter $\alpha$ increases. Higher states 
$(n=2)$ rise faster than the ground state $(n=0)$, 
since energy and mass are related, an increase in 
$\alpha$ leads to a higher in our calculated mass.
	\subsection{Heavy Quarkonia}
In this section, we calculate the masses of heavy quarkonia in three dimensions employing the same relation as in Ref. \cite{s};
\begin{equation}
	M= m_{1}+ m_{2}+ E_{\nu,\gamma}\label{x26},
\end{equation}
where $E_{\nu,\gamma}$ is given from Eq. (\ref{x24}). 
The potential parameters \textit{a, b}, and \textit{d} are calculated by fitting experimental data and listed in Table (\ref{tab:t1}). We use the quark masses of $m_{c} = 1.533$ GeV and $ m_{b} = 4.666$ GeV.
\subsubsection{The Effect of Topological Defect}
\begin{figure}[!h]
	\begin{minipage}{.48\textwidth}
		\centering
		\includegraphics[width=1.05\linewidth]{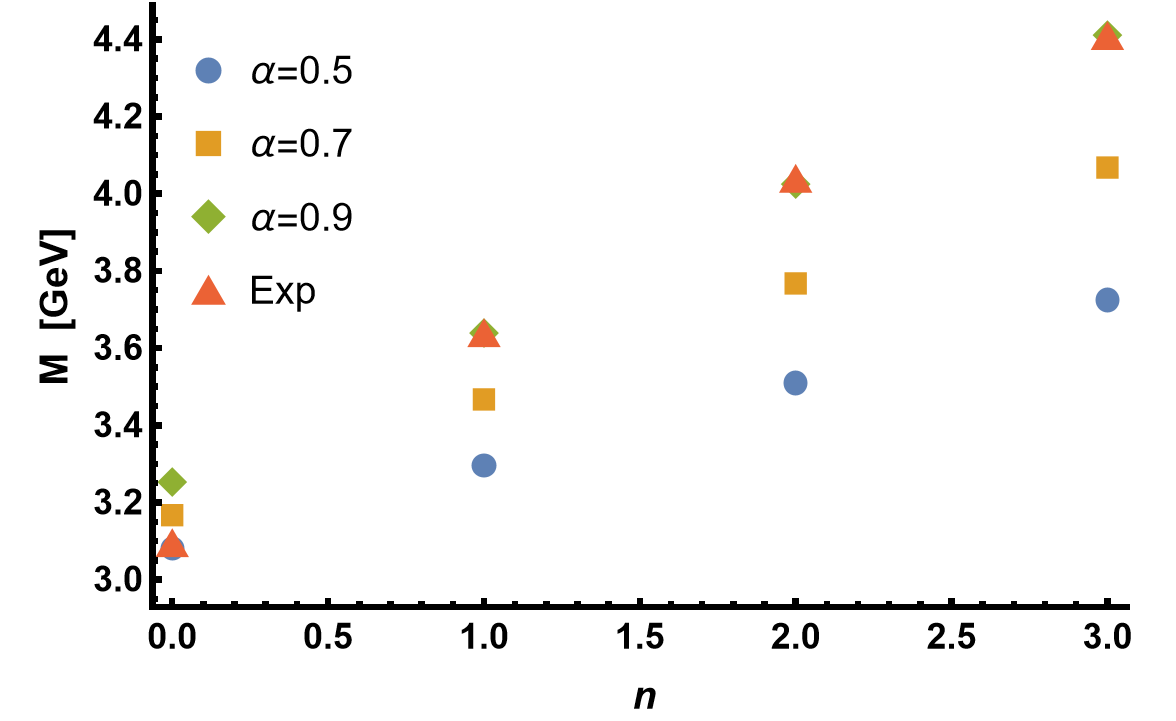}
		\textbf{(a) Charmoniunm}
	\end{minipage}
	\quad	
	\begin{minipage}{0.48\textwidth}
		\centering
		\includegraphics[width=1.05\linewidth]{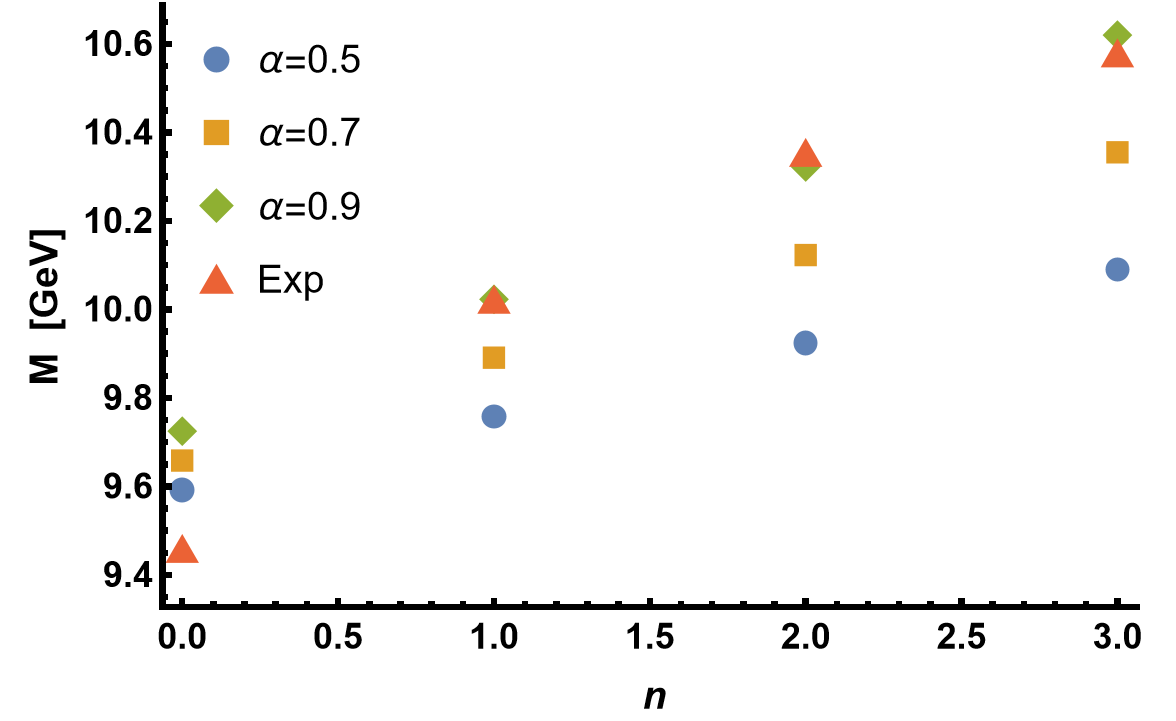} 
		\textbf{	(b) Bottomnium}
	\end{minipage}
	\\
	\caption{The $S$-states mass spectra for charmonium and bottomonium at different topological defect, }\label{h3}
\end{figure}

Table (\ref{tab:t2}) shows the mass spectrum of charmonium in various states, assuming topological defects  $\alpha=0.7$ and $\alpha=0.9$, and at the amount of magnetic flux $\Phi=0.008$. The present results have been improved over those stated in the recent Refs. \cite{s}-\cite{c81} and show good agreement with experimental findings.
In case $\alpha=0.9$, the values of state $1P$, $2S$, $2P$, and $4S$ are near-perfect matches to the experimental data. In the present work, the topological defect takes values $0 < \alpha  <  1$. Additionally, we have computed the total error for all states = $\sum_{states}|\frac{M_{Our}-M_{Exp.}}{n M_{Exp.}}|$. 
where $M_{Our}$ is our estimated mass and $M_{Exp.}$ is the experimental mass and, $n$ is the number of states. To compare with experimental data this method provides high accuracy compared to previous research. 
In Table (\ref{tab:t2}), we give the overall error for the charmonium mass as $0.042\%$ in the first case and $0.013\%$ in the second case, both of which are lower than the total error reported in prior studies.
In Ref. \cite{s}, the author examined the conformable fractional analytical-exact iteration approach to derive the analytical solutions of the N-dimensional radial Schrödinger equation, calculating the mass spectra of heavy quarkonia, resulting in a total error of $0.033\%$ for this work.
In Ref. \cite{nor5}, the Schrödinger equation is analytically resolved using the generalised fractional extension of the Nikiforov-Uvarov method, resulting in the mass spectra of heavy quarkonia; hence, the overall error of this study is  $0.023\%$. 
In Ref. \cite{s15}, the Trigonometric Rosen-Morse is used to calculate heavy-meson properties in the free and hot media. We calculated their total error of heavy meson is $0.024\%$.
The Schrödinger equation is resolved in Ref. \cite{s20} using the asymptotic iteration method with Cornell potential, resulting in a total error of $0.05\%$. In Ref.  \cite{c81}, the non-relativistic approach using a phenomenological QCD model to determine the diquark masses and tetraquark $cc\bar{c}\bar{c}$ .\\
In Table (\ref{tab:t3}), the mass spectra of bottomonium are generated using the topological defects $\alpha=0.7$ and $\alpha=0.9$. 
As can be seen, our results are superior to those of the previous Refs. \cite{s}-\cite{c100} are consistent with the experimental findings. 
The total error for bottomonium mass was calculated to be $0.016\%$ and $0.006\%$, as presented in Table (\ref{tab:t3}), indicating a reduction in total error relative to previous studies.
In Ref. \cite{c100}, the mass spectra are computed by the author using the non-relativistic potential model, the overall error of this study is  $0.003\%$.
Fig. (\ref{h3}) presents the $S-$state of the mass spectra for both charmonium and bottomonium. These values were obtained using Eqs. (\ref{x24}) and (\ref{x26}) across a range of topological defect parameters ($\alpha$).
The results demonstrate that as $\alpha$ increases, the predicted masses converge significantly with experimental values, outperforming the classical scenario ($\alpha=1$). This correlation underscores the essential role topological defects play in accurately modeling these mass spectra.
\subsubsection{The Effect of Magnetic Flux}
This variant shows how the masses of charmonium and bottomonium are affected by $\Phi$. While keeping $\alpha$ at $0.9$ in Table (\ref{tab:t4}), we investigate the mass of charmonium in the cases where$\Phi=0.1$, $\Phi=0.5$, and $\Phi=0.7$. When compared to earlier studies, the overall error are $0.05\%$, $0.02\%$, and $0.013\%$, indicating that the newly added parameters maintain the physical consistency of the projected mass spectra. In Table (\ref{tab:t5}). we analyse the mass of bottomonium when $\alpha$ is maintained at $0.9$ and $\Phi=0.1$, $\Phi=0.5$, and $\Phi=0.7$. The total errors are $0.021\%$, $0.014\%$, and $0.008\%$, and it demonstrates that, in comparison to earlier research, the newly added parameters preserve the physical consistency of the projected mass spectra.
	\begin{table}[h!]
	\caption{ parameters of potential.}
	\begin{center}
		\begin{tabular}{cccc}
			\hline
			&$ a(GeV^{3})$ &  $b (GeV^{2})$ &  $ d (GeV^{-1})$ \\
			\hline
			$c\bar c$&0.0705&0.4498&0.8361\\
			$b\bar b$&0.1283&0.4753&0.5075\\
			\hline
			\hline
		\end{tabular}
		\centering
		\label{tab:t1}
	\end{center}	
\end{table}
\begin{table}[!h]
	\caption{\small{Charmonium mass spectra at $\Phi=0.008$, we take $\alpha=0.7$  and $\alpha=0.9$ with other  studies in $(GeV)$}.}\label{tab:t2}
	\small
	\renewcommand{\arraystretch}{1.1}
	\begin{center}
		\begin{tabular}{c|cccccccc}
			\hline
			State&$ M(\alpha=0.7$)&$M(\alpha=0.9)$&Ref.\cite{s}&Ref.\cite{nor5}&Ref.\cite{s15}& Ref.\cite{s20}&  Ref.\cite{c81}& Exp.\cite{s32}\\
			\hline
			$1S$&3.167&3.253&3.096&3.096&3.239&3.096&3.0067&3.096\\
			
			$1P$&3.428&3.514&3.344&3.570&3.372&3.214&3.256&3.520\\
			
			$2S$&3.467&3.639&3.786&3.580&3.646&3.412&3.676&3.640\\
			
			$1D$&3.795&3.881&3.769&4.050&3.604&3.686&3.556&3.760\\
			
			$2P$&3.728&3.900&4.034&4.040&3.779&3.773&3.98&3.900\\
			
			$3S$&3.768&4.025&4.270&4.040&4.052&4.275&--&4.040\\
			
			$4S$&4.068&4.411&4.621&4.510&4.459&4.865&--&4.410\\
			
			Totatl error&0.042&0.013&0.033&0.023&0.024&0.050&--&-- \\
			\hline
			\hline
		\end{tabular}
	\end{center}	
\end{table}			
\begin{table}[!h]
	\caption{\small{Bottomonium mass spectra at $\Phi=0.012$, we take $\alpha=0.7$  and $\alpha=0.9$ with other studies in $(GeV)$}.}\label{tab:t3}
	\small
	\renewcommand{\arraystretch}{1.1}
	\begin{center}
		\begin{tabular}{c|cccccccc}
			\hline
			State&$M(\alpha=0.7)$ &$M(\alpha=0.9)$&Ref.\cite{s}& Ref.\cite{nor5}&Ref.\cite{s15}&Ref.\cite{s20}& Ref.\cite{c100}& Exp.\cite{s32}\\
			\hline
			$1S$&9.659&9.725&9.444&9.460&9.495&9.460& 9.409&9.460\\
			
			$1P$&9.833&9.900&9.711&9.850&9.657&9.492&9.863&9.900\\
			
			$2S$&9.891&10.023&9.946&9.840&10.023&10.023&9.987&10.023\\
			
			$1D$&10.093&10.160&10.161&10.230&10.161&9.551&10.135&10.160\\
			
			$2P$&10.065&10.198&10.213&10.220&10.260&10.038&10.213&10.266\\
			
			$3S$&10.132&10.322&10.447&10.210&10.355&10.585&10.325&10.355\\
			
			$4S$&10.355&10.620&10.949&10.580&10.579&11.148&10.596&10.580\\
			
			Total error&0.016&0.006&0.011&0.007&0.004& 0.028& 0.003&-- \\
			\hline
			\hline
		\end{tabular}
	\end{center}	
\end{table}
\begin{table}[!h]
	\caption{\small{Charmonium mass spectra at  $\alpha=0.9$ and we take $\Phi=0.008$, $\Phi=0.1$, $\Phi=0.5$, and $\Phi=0.7$  with other  studies in $(GeV)$}.}\label{tab:t4}
	\small
	\renewcommand{\arraystretch}{1.1}
	\begin{center}
		\begin{tabular}{c|cccccccccc}
			\hline
			State&$\Phi=0.008$&$\Phi=0.1$&$\Phi=0.5$&$\Phi=0.7$&\cite{s}&\cite{nor5}&\cite{s15}& \cite{s20}&\cite{c81}& Exp.\cite{s32}\\
			\hline
			$1S$&3.253&3.25&3.22&3.24&3.096&3.096&3.239&3.096&3.0067&3.096\\
			
			$1P$&3.514&3.507&3.38&3.47&3.344&3.570&3.372&3.214&3.256&3.520\\
			
			$2S$&3.639&3.64&3.606&3.62&3.786&3.580&3.646&3.412&3.676&3.640\\
			
			$1D$&3.881&3.87&3.72&3.83&3.769&4.050&3.604&3.686&3.556&3.760\\
			
			$2P$&3.900&3.89&3.77&3.86&4.034&4.040&3.779&3.773&3.98&3.900\\
			
			$3S$&4.025&4.02&3.99&4.009&4.270&4.040&4.052&4.275&--&4.040\\
			
			$4S$&4.411&4.408&4.378&4.39&4.621&4.510&4.459&4.865&--&4.410\\
			
			Totatl error&0.042&0.05&0.02&0.013&0.033&0.023&0.024&0.050&--&-- \\
			\hline
			\hline
		\end{tabular}
	\end{center}	
\end{table}	

\begin{table}[!h]
	\caption{\small{Bottomonium mass spectra at$\alpha=0.9$ and we take $\Phi=0.012$, $\Phi=0.1$, $\Phi=0.5$, and $\Phi=0.7$ with other studies in $(GeV)$.}}\label{tab:t5}
	\small
	\renewcommand{\arraystretch}{1.1}
	\begin{center}
		\begin{tabular}{c|cccccccccc}
			\hline
			State&$\Phi=0.012$ &$\Phi=0.1$&$\Phi=0.5$&$\Phi=0.7$&\cite{s}& \cite{nor5}&\cite{s15}&\cite{s20}& \cite{c100}&Exp. \cite{s32}\\
			\hline
			$1S$&9.725&9.637&9.699&9.716&9.444&9.460&9.495&9.460& 9.409&9.460\\
			
			$1P$&9.900&9.698&9.798&9.879&9.711&9.850&9.657&9.492&9.863&9.900\\
			
			$2S$&10.023&9.869&9.998&10.15&9.946&9.840&10.023&10.023&9.987&10.023\\
			
			$1D$&10.160&9.908&10.026&10.135&10.161&10.230&10.161&9.551&10.135&10.160\\
			
			$2P$&10.198&9.9304&10.0966&10.178&10.213&10.220&10.260&10.038&10.213&10.266\\
			
			$3S$&10.322&10.102&10.297&10.314&10.447&10.210&10.355&10.585&10.325&10.355\\
			
			$4S$&10.620&10.334&10.592&10.612&10.949&10.580&10.579&11.148&10.596&10.580\\
			
			Total error&0.006&0.021&0.014&0.008&0.011&0.007&0.004& 0.028& 0.003&-- \\
			\hline
			\hline
		\end{tabular}
	\end{center}	
\end{table}

\begin{table}[!h]
	\caption{The masses of tetraquark (in GeV) and comparison with other works.}\label{tab:t6}
	\small
	\centering
	
	\renewcommand{\arraystretch}{1.3}
	\begin{center}
		\begin{tabular}{c|c|cccccc}
			\hline
			Tetraquark& Mass &$1S$&$1P$& $2S$&$2P$&$3S$&$4S$\\
			\hline
			\multirow{6}*{$cc\bar{c}\bar{c}$} &$M(\alpha=0.7,\Phi=0.008)$&6.327&6.470&6.536&6.679&6.745&6.954\\
			
			&$M(\alpha=09,\Phi=0.008)$&6.496&6.636&6.702&6.842&6.908&7.114\\
			&$M(\alpha=0.9,\Phi=0.1)$&6.543&6.593&6.808&6.858&7.072&7.337\\
			&$M(\alpha=0.9,\Phi=0.5)$&6.553&6.690&6.818&6.955&7.084&7.348\\
			&$M(\alpha=0.9,\Phi=0.7)$&6.539&6.621&6.804&6.886&7.069&7.334\\
			& Ref.\cite{nor5}& 5.930&6.204& 6.202 &6.480&6.470&6.750\\ 
			&Ref.\cite{c81}&6.909&6.999&7.073&7.143&--&--\\
			&Ref.\cite{ph}& 6.198& --&--&--&--&--\\
			
			&Ref.\cite{n34}&6.354&--&--&--&--&--\\
			&Ref.\cite{nor4}&6.609&6.611&--&--&--&--\\
			&Ref.\cite{cc}&6.19&6.631&6.782&7.091&7.259&--\\
			&Ref.\cite{c79}&6.498&6.74&7.007&--&7.024&--\\
			\cline{2-8}
			\hline
			\hline
			\multirow{6}*{$bb\bar{b}\bar{b}$}&$M(\alpha=0.7,\Phi=0.012)$&19.561&19.653&19.722&19.814&19.884&20.045\\
			
			&$M(\alpha=0.9,\Phi=0.012)$&19.739&19.831&19.946&20.037&20.152&20.359\\
			&$M(\alpha=0.9,\Phi=0.1)$&19.682&19.712&19.842&19.873&20.003&20.164\\
			&$M(\alpha=0.9,\Phi=0.5)$&19.688&19.773&19.849&19.933&20.009&20.17\\
			&$M(\alpha=0.9,\Phi=0.7)$&19.68&19.729&19.841&19.89&20.002&20.162\\
			&Ref.\cite{nor5}&18.650&18.930&18.920&19.190&19.180&19.450\\
			&Ref.\cite{c100}&18.719&19.381&19.791&--&19.479&--\\
			&Ref.\cite{ph}&18.754&--&--&--&--&--\\
			
			&Ref.\cite{n34}&19.673&--&--&--&--&--\\
			
			&Ref.\cite{nor4}&20.012&20.0116&--&--&--&--\\
			&Ref.\cite{cc}&19.315&19.536&19.680&19.820&19.941&--\\
			&Ref.\cite{c64}&18.826&19.281&19.335&19.597&19.644&--\\
			
			\cline{2-8}
			\hline
			\hline
		\end{tabular}
	\end{center}	
\end{table}
	\subsection{Tetraquark}
The tetraquark is a two-body non-relativistic system comprising a diquark and an antidiquark that interacts via a Cornell potential, facilitating accurate predictions of the tetraquark as demonstrated in Refs. \cite{n43, ph}.
We propose to establish a four-particle, two-body diquark-antidiquark system to investigate diquark states within the tetraquark framework. The theory of tetraquarks as diquark-antidiquark configurations is employed to formulate a technique for finding the quantitative masses of certain tetraquark states. 
To describe the Hamiltonian of a quark-antiquark system, an unperturbed one-gluon exchange (OGE) potential is used, along with a perturbation component to account for spin. 
Subsequently, by calculating diquark masses and assuming the same mass for the corresponding antidiquark systems due to charge conjugation symmetry, the model is expanded to include diquark-antidiquark systems \cite{ph}.
\newpage
Table (\ref{tab:t6}) shows our estimated tetraquark masses,
it show the mass of the tetraquark $cc\bar{c}\bar{c}$ increases from $6.327$ GeV to $6.954$ GeV at state $1S$ in case $\alpha=0.7$,$\Phi=0.008$, and also shows increasing from $6.496$ GeV to $6.7.114$ GeV in case $\alpha=0.9$,$\Phi=0.008$.\\
The tetraquark  $bb\bar{b}\bar{b}$ exhibits an analogous pattern, with its mass increasing from $19.682$ GeV to $20.164$ GeV at a state $1S$ in case $\alpha=0.9$,$\Phi=0.1$, and  also increasing from $19.688$ GeV to $20.17$ GeV in case $\alpha=0.9$,$\Phi=0.5$ 
Additionally, the results show that relatively small variations in the projected masses are produced by changes in the model parameters $\alpha$ and $\Phi$. The reliability of the model predictions is increased by this behaviour, which shows that the produced spectra are comparatively stable against parameter fluctuations.
\clearpage
Comparison with previous theoretical studies shows that the present predictions fall within the range of values reported in the literature. For the $cc\bar{c}\bar{c}$ state, the obtained ground-state masses are generally closer to the predictions of Refs. \cite{n34}, \cite{nor4}, and \cite{c79}, while remaining higher than those reported in Ref. \cite{nor5}. For the $bb\bar{b}\bar{b}$ tetraquark, the present results are in reasonable agreement with Refs. \cite{n34, nor4}. Such differences may originate from the use of different potential models, assumptions regarding quark interactions, and relativistic corrections.
In Ref. \cite{ph}, the Schrödinger equation is numerically solved using the Cornell potential with the spin-spin interactions. The solution to the Schrödinger equation is obtained by studying the non-relativistic Hulthen potential, linear confining potential, and spin-spin interaction in \cite{n34}. In Ref. \cite{nor4}, the author utilised the non-relativistic Bethe-Salpeter equation to explain a tetraquark, using the logarithmic potential, the harmonic potential, and the linear potential with the spin-spin interactions. In Ref. \cite{cc}, masses of the fully-heavy tetraquarks are calculated in the diquark-antidiquark picture within the relativistic quark model based on the quasipotential approach and quantum chromodynamics. In Ref. \cite{c79}, using a nonrelativistic quark model, the author computed the mass spectrum of the higher excited fully charmed tetraquark $cc\bar{c}\bar{c}$ states, including the s-wave radial excitations and the \textit{P}-wave states.
In Ref. \cite{c64}, using a relativised diquark model Hamiltonian, the relativistic quark model, which depends on the quasi-potential technique and quantum chromodynamics, utilises the fully-heavy tetraquarks in the diquark-antidiquark picture.
Overall, the agreement with previous works and the smooth behaviour of the mass spectra support the validity of the adopted theoretical framework for describing fully heavy tetraquark systems.
\section{Conclusion}
Exploring the interplay of topological defects and external magnetic fields on heavy quarkonia and fully-heavy tetraquarks is vital for understanding how environmental and geometrical alterations alter fundamental hadronic characteristics and color confinement mechanisms.
In this work, we investigated the analytical solutions of the non-relativistic Schrödinger equation for heavy quarkonia and tetraquark systems. The study incorporates a comprehensive potential model that combines the Cornell, harmonic, and inverse-harmonic terms under the influence of topological defects $(\alpha)$ and external magnetic flux $(\Phi)$. The effective potential analysis confirms that the model successfully captures both short-range repulsion and long-range color confinement, which are essential for the formation of stable hadronic bound states. The energy eigenvalues (E) were found to increase linearly with the topological defect parameter $(\alpha)$. This demonstrates that the geometric structure of the background spacetime directly shifts the energy levels of the system. Based on the calculated energy levels, we determined the mass spectra for various diquark states; these were then utilised as building blocks to estimate the masses of heavy tetraquark systems. The numerical data in our tables indicate that mass increases with the principal and orbital quantum numbers $(n, l)$. We compared our results for charmonium, bottomonium, and tetraquark masses with experimental data and other theoretical models. The results show a high level of agreement, particularly for heavy quark states, which validates the reliability of our proposed potential and the analytical method used. The interactions between the confinement potential and the topological constraints provide new insights into the stability of exotic states.
The obtained results show that the addition of the magnetic flux parameter $\Phi$ significantly influences the mass spectra of heavy quarkonia and heavy tetraquark states. Changes in the magnetic flux were found to cause observable shifts in the computed masses, suggesting that it has an impact on the system's energy levels and confinement dynamics. Additionally, the addition of magnetic flux gives the model greater flexibility and improves the way quarkonia and the tetraquark spectra are described These results imply that the spectroscopic features of exotic hadronic systems can be greatly altered by external magnetic forces, which should be considered while researching these systems.
We hope to extend this work to spin-dependent interactions (hyperfine and fine structure) as future work by using the Draic equation as in Ref. \cite{c55}.

\end{document}